\documentclass[showpacs,aps,amsmath,amssymb,twocolumn]{revtex4}
\usepackage{CJK}

\usepackage[latin1]{inputenc}
\usepackage{graphicx}
\usepackage{bm}
\usepackage{xcolor}
\usepackage{array}
\usepackage{braket}
\usepackage{mathtools}
\usepackage{microtype}
\usepackage{wasysym}
\renewcommand{\baselinestretch}{1.0}
\begin{document}

\renewcommand{\baselinestretch}{1.0}
\title{Influence of polarization and the environment on wave-particle duality}

\author{Andr\'{e}a Freire dos Santos, Nat\'{a}lia E. L. Barbosa, J. L. Montenegro Ferreira,\\ and Bert\'{u}lio de Lima Bernardo}

\email{bertulio.fisica@gmail.com}

\affiliation{Departamento de F\'{\i}sica, Universidade Federal da Para\'{\i}ba, 58051-900 Jo\~ao Pessoa, PB, Brazil}

\begin{abstract}
  
Wave-particle duality is certainly one of the most curious concepts of contemporary physics, which ascribes mutually exclusive behaviors to quantum systems that cannot be observed simultaneously. In the context of two-path interferometers, these two behaviors are usually described in terms of the visibility of interference fringes and the path distinguishability. Here, we use quantum information-theoretic tools to derive quantifiers of these two properties, which accounts for the combined influence of path probability and polarization, and demonstrate that they satisfy a complementarity relation. We further show that the derived quantities can work as probes in the study of open quantum dynamics by revealing interesting facets of environment actions, such as: decoherence, depolarization and scattering.                                                  
   
\end{abstract}

\maketitle


\section{Introduction}

One of the most curious and intriguing features of quantum mechanics is the wave-particle behavior. In 1928, Bohr called attention to this dual aspect of quantum systems, which sometimes manifest themselves as waves and sometimes as particles, depending on the experimental setup to which they are submitted \cite{bohr}. In the context of the double-slit experiment, the complementary wave and particle behaviors were pointed out by Feynman as ``the only mystery'' of quantum mechanics \cite{feynman}. Later on, a quantitative approach to this problem was presented by Wootters and Zurek, who showed that the two complementary aspects might be simultaneously present \cite{wootters}. 

Such ideas were followed by many subsequent works until Englert, Greenberger and Yasin proposed a simple complementarity relation expressing the wave behavior in terms of the visibility $\mathcal{V}$ of the interference fringes produced in the experiment, and the particle behavior in terms of the distinguishability $\mathcal{D}$ between the two possible paths taken by, e.g., photons in a interferometer. This is known as the EGY inequality \cite{englert,greenberger}:
\begin{eqnarray} 
\label{1}
\mathcal{D}^{2} + \mathcal{V}^{2} \leq 1.
\end{eqnarray}
Here, we call attention to the fact that the influence of internal degrees of freedom of the particles, such as spin or polarization, are not considered in deriving $\mathcal{V}$ and $\mathcal{D}$, i.e., a beam of polarized particles is assumed as a hypothesis. 

In this relation, $\mathcal{V}=1$ and $\mathcal{D} = 0$ mean maximum visibility of the interference fringes, and no knowledge about which path the particles take. This is the case of complete wave-like behavior. When $\mathcal{V} = 0$ and $\mathcal{D}=1$, we have absence of interference pattern, and information about the path of the particles is totally available. Still, there are intermediate situations, in which $\mathcal{D}\neq0$ and $\mathcal{V}\neq0$, that indicate partial which-path information and interference fringes with limited contrast. Experimental tests of the complementary behaviors have been realized with atoms \cite{durr}, nuclear magnetic resonance setups \cite{roy, auccaise}, and photons \cite{jacques,brien}.

Interestingly, studies of the Bohr's complementarity principle have recently gained renewed interest owing in part to experimental advances in photonic quantum technologies \cite{brien}. It has been demonstrated that the degree of polarization of the photons emerging from the source is directly connected to the manifestation of the complementary effects in a two-path interferometer \cite{lahiri,zela,eberly,norrman2}. The purity of the source has been theoretically and experimentally demonstrated to be closely related to the entanglement between the path and the remaining degrees of freedom of the particles \cite{qian,yoon}. In parallel, generalizations of these developments to the framework of multipath interferometers have also brought attention \cite{brito,qureshi}.

In this work, we provide a different approach to describe the wave-particle duality in the double-slit scheme, where the polarization degree of freedom is considered as a fundamental property to completely define the concepts of visibility and distinguishability. Namely, we derive quantifiers of visibility and distinguishability, denoted respectively by $V$ and $D$, that embodies both the influence of the probability of each path and the polarization, and demonstrate that they satisfy a complementarity relation analogous to the EGY inequality. Additionally, we propose the use of these quantities to probe system-environment interactions. We observe that effects such as decoherence, depolarization and scattering leave observable imprints in the time evolution of $V$ and $D$. 

In the following section we present a density matrix description of coherence and polarization, which is the approach we take here. In Sec. III we derive the quantifiers of visibility and distinguishability. In Sec. IV we demonstrate the  complementarity relation involving the derived quantifiers. Sec. V presents some key examples that help understanding their physical meaning. In Sec. VI we investigate how they can be useful in the study of open quantum systems. Sec. VII then summarizes our findings and points out some perspectives.

\section{Preliminaries}

\begin{figure}[ht]
\centerline{\includegraphics[width=7.5cm]{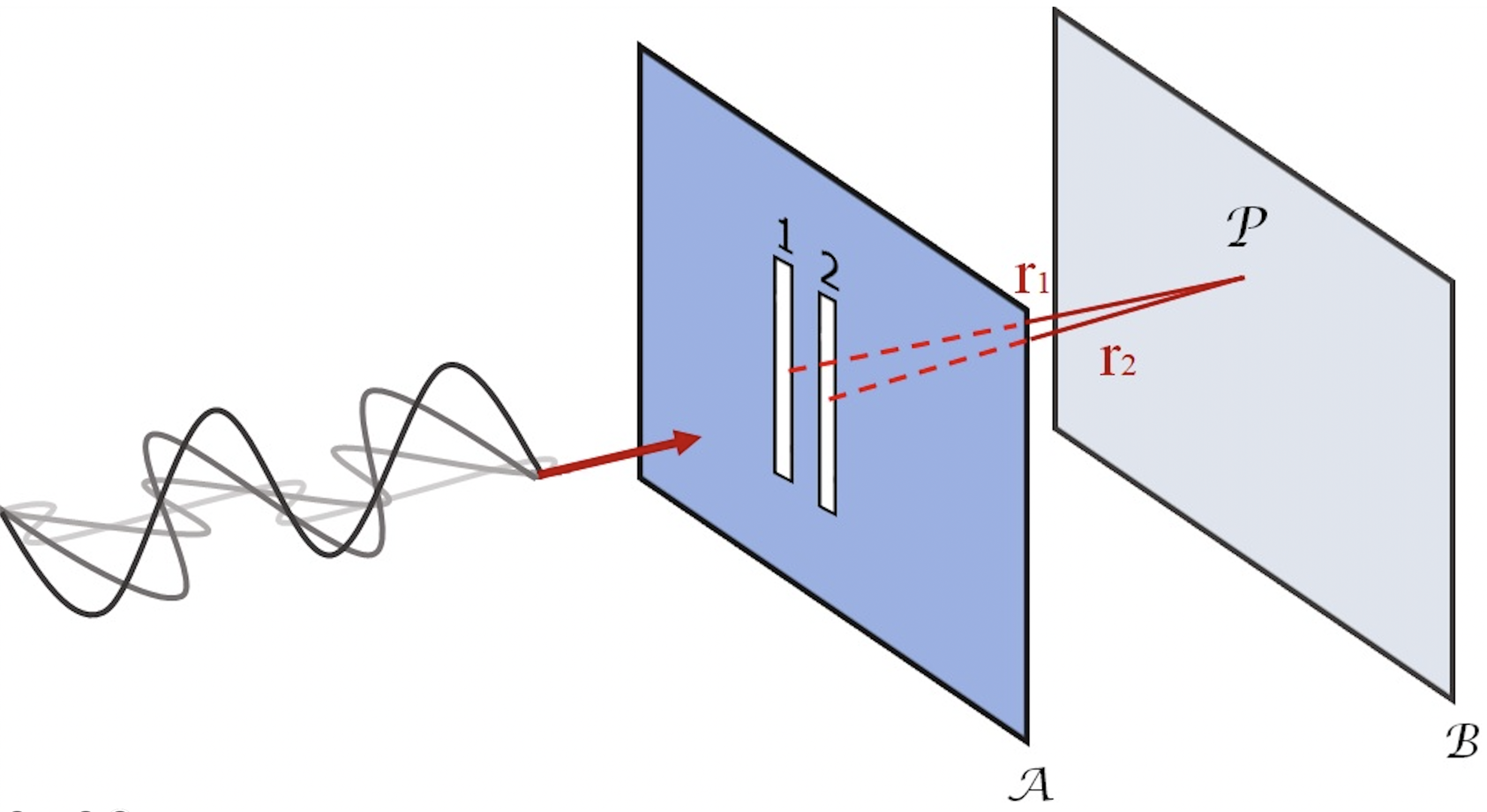}}
\caption{(Color online) Double-slit experiment. Photons traverse a mask $\mathcal{A}$ through the slits $1$ and $2$ with an arbitrary polarization state to have their position marked at some point $\mathcal{P}$ on the detection screen $\mathcal{B}$.}
\label{setup}
\end{figure}

In this section we present the density matrix formalism that describes the coherence and polarization properties of light in the double-slit scenario developed in Ref. \cite{bernardo}, whose classical version was previously proposed by Wolf \cite{wolf}. Consider an ensemble of photons submitted to a double-slit apparatus $\mathcal{A}$, whose position are marked on a distant detection screen $\mathcal{B}$, as shown in Fig. $\ref{setup}$. The photons arrive at a given point $\mathcal{P}$ with a probability that depends on: i) whether they pass through slit 1 or 2, whose path states we denote respectively by $\ket{1}$ and $\ket{2}$, and ii) the polarization state, here spanned by the horizontal  $|H\rangle$ and vertical $|V\rangle$ basis states. These two conditions are assumed because the formation of interference fringes is both a path- and polarization-dependent phenomenon.

In order to provide a quantum information-theoretic description of such an experiment, we introduce the coherence-polarization density matrix in the form
\begin{equation}
\label{2}
\hat{\rho} = 
\begin{pmatrix}
\rho_{11} & \rho_{12} & \rho_{13} & \rho_{14} \\
\rho_{21} & \rho_{22} & \rho_{23} & \rho_{24} \\
\rho_{31} & \rho_{32} & \rho_{33} & \rho_{34} \\
\rho_{41} & \rho_{42} & \rho_{43} & \rho_{44}
\end{pmatrix},
\end{equation}  
which is represented here in the basis $\{ \ket{H,1}, \ket{H,2}, \ket{V,1}, \ket{V,2} \}$. To clarify the notation, we have for example that the basis state $\ket{H,1}$ represents the state of a horizontally-polarized photon that passes through slit 1, and so on. The density matrix of Eq. (\ref{2}) can be used to calculate the probability density $\rho (\mathcal{P})$ of detecting a photon at any point $\mathcal{P}$ on the screen. We do so by using the fact that horizontally-polarized photons do not interfere with vertically-polarized ones, which
means that the probability density splits into two independent contributions as
follows
\begin{equation} \label{4}
\rho (\mathcal{P}) = \braket{H,\mathcal{P}|\hat{\rho}|H,\mathcal{P}} + \braket{V,\mathcal{P}|\hat{\rho}|V,\mathcal{P}},
\end{equation}
where $\ket{H,\mathcal{P}}$ ($\ket{V,\mathcal{P}}$) represents the state of a photon at $\mathcal{P}$ with horizontal (vertical) polarization.

Now, in considering that the wavelength of the photons is of the order of the size of the apertures, we can assume the diffraction limit. In this case, the spatial probability amplitudes of the photons that
emerge from the slits can be described by spherical waves. Namely,
\begin{eqnarray} \label{5}
     \langle{H,1|H,\mathcal{P}}\rangle = \langle{V,1|V,\mathcal{P}}\rangle = \frac{e^{ikr_{1}}}{r_{1}},
\end{eqnarray}
\begin{eqnarray} \label{6}
     \langle{H,2|H,\mathcal{P}}\rangle = \langle{V,2|V,\mathcal{P}}\rangle = \frac{e^{ikr_{2}}}{r_{2}}.
\end{eqnarray}
By substitution of Eqs. (\ref{5}) and (\ref{6}) into Eq. (\ref{4}), we obtain that the probability density of detecting a photon at $\mathcal{P}$ is
\begin{eqnarray} \label{7}
\rho(\mathcal{P})&=& \rho_{1}(\mathcal{P}) + \rho_{2}(\mathcal{P}) \nonumber \\
&+& 2\sqrt{\rho_{1}(\mathcal{P})} \sqrt{\rho_{2}(\mathcal{P})}Re[\mu e^{ik(r_{1} - r_{2})}],
\end{eqnarray}
where $\rho_{1}(\mathcal{P}) = (\rho_{11}+\rho_{33})/ r^{2}_{1}$ and $\rho_{2}(\mathcal{P}) = (\rho_{22}+\rho_{44})/ r^{2}_{2}$ are the individual probability densities of detecting a photon at $\mathcal{P}$ when they emerge exclusively from slit $1$ and $2$, respectively. We also have that $Re$ denotes the real part, and the definition of degree of coherence is given by
\begin{eqnarray}\label{8}
\mu = \frac{\rho_{12} + \rho_{34}}{\sqrt{\rho_{11} + \rho_{33}}\sqrt{\rho_{22} + \rho_{44}}},
\end{eqnarray}
which  satisfies the relation $0\leq |\mu| \leq 1$. The cases in which $\mu=0$ and $|\mu| = 1$ mean incoherent and coherent photons, respectively, and when $0 < |\mu| < 1$ we have partially coherent photons.

Now we turn our attention to the study of the polarization. Similar to electromagnetic optics, the polarization state of the photons can be characterized by the Stokes parameters, which in the present context are given by the ensemble average of the (polarization) Pauli operators associated with each path \cite{abouraddy,james}.  As such, the polarization state of the photons that travel through slit 1 is characterized by the Stokes parameters  \cite{bernardo}:
\begin{equation}
\label{s01}
S_{0}^{(1)}=Tr[(|H,1\rangle\langle H,1| + |V,1\rangle\langle V,1| )\hat{\rho}] = \rho_{11} + \rho_{33},
\end{equation}
   \begin{equation}
   S_{1}^{(1)}=Tr[(|H,1\rangle\langle H,1| - |V,1\rangle\langle V,1| )\hat{\rho}] = \rho_{11} - \rho_{33},
    \end{equation}
    \begin{equation}
   S_{2}^{(1)}=Tr[(|H,1\rangle\langle V,1| + |V,1\rangle\langle H,1| )\hat{\rho}] = \rho_{13} + \rho_{31},
   \end{equation}
   \begin{equation}
   S_{3}^{(1)}= -i\left\{Tr[(|H,1\rangle\langle V,1| - |V,1\rangle\langle H,1| )\hat{\rho}]\right\} = i(\rho_{13} - \rho_{31}).
\end{equation}
Similarly, the corresponding Stokes parameters of the photons passing through slit $2$ are given by
\begin{equation}
 S_{0}^{(2)}=Tr[(|H,2\rangle\langle H,2| + |V,2\rangle\langle V,2| )\hat{\rho}] = \rho_{22} + \rho_{44},
 \end{equation}
 \begin{equation}
 S_{1}^{(2)}=Tr[(|H,2\rangle\langle H,2| - |V,2\rangle\langle V,2| )\hat{\rho}] = \rho_{22} - \rho_{44},
  \end{equation}
  \begin{equation}
 S_{2}^{(2)}=Tr[(|H,2\rangle\langle V,2| + |V,2\rangle\langle H,2| )\hat{\rho}] = \rho_{24} + \rho_{42},
 \end{equation}
 \begin{equation}
 \label{s23}
 S_{3}^{(2)}= -i\left\{Tr[(|H,2\rangle\langle V,2| - |V,2\rangle\langle H,2| )\hat{\rho}]\right\} = i(\rho_{24} - \rho_{42}).
\end{equation}

Having established these parameters, they can be used to define the degree of polarization of the photons that take path $j$ according to the formula 
$p_{j} = \sqrt{(S_{1}^{(j)})^{2} + (S_{2}^{(j)})^{2} + (S_{3}^{(j)})^{2}}/S_{0}^{(j)}$, with $j = 1,2$. After some calculation, we find the degrees of polarization associated with paths 1 and 2 respectively as
\begin{eqnarray}\label{9}
   p_{1} &=& \sqrt{1 - \frac{4(\rho_{11}\rho_{33} - \rho_{13}\rho_{31})}{(\rho_{11} +\rho_{33})^{2}}},
\end{eqnarray}
\begin{eqnarray}\label{10}
    p_{2} &=&\sqrt{1 - \frac{4(\rho_{22}\rho_{44} - \rho_{24}\rho_{42})}{(\rho_{22} +\rho_{44})^{2}}}.
\end{eqnarray}
Here, $p_{j}=0$ and $p_{j} = 1$ signify unpolarized and polarized photons, respectively. The cases in which $0 < p_{j} < 1$ correspond to partial polarization. The elements of the density matrix obey the properties $\rho_{nn}\rho_{mm} \geq  |\rho_{nm}|^2$ and $\rho^{2}_{nn} + \rho^{2}_{mm} \geq 2 \rho_{nn}\rho_{mm}$ \cite{neumann}, which guarantees that $p_{j} \in [0,1]$.

In the end, as can be seen from Eqs.~(\ref{8}), (\ref{9}) and (\ref{10}), the coherence-polarization density matrix $\hat{\rho}$ of Eq. (\ref{2}) contains all information about the coherence and polarization of the ensemble of photons.

\section{Visibility and Distinguishability}

Visibility is a measure of the contrast of the interference fringes that appear on the detection screen, being therefore expressed in terms of the maximum and minimum intensities of the fringes \cite{mandel}. The coherence-polarization density matrix can provide this information through the relation
\begin{eqnarray}
\label{11}
V = \frac{\rho_{max} - \rho_{min}}{\rho_{max} + \rho_{min}},
\end{eqnarray}
where $\rho_{max}$ and $\rho_{min}$ are the maximum and minimum values of the probability density of Eq.~(\ref{7}), which are obtained when the third term in the right-hand side changes as a consequence of the variation of $r_1$ and $r_2$. Since $\mu$ is a complex number, we can write $\mu = |\mu|e^{i\varphi}$, which allows us to rewrite Eq.~(\ref{7}) as 
\begin{equation}
\rho = \rho_{1} + \rho_{2} + 2\sqrt{\rho_{1} \rho_{2}}
|\mu| \cos[k(r_1 - r_2) + \varphi],
\end{equation}
where we omitted the dependence on $\mathcal{P}$. In this form, we have that $\rho_{max/min} = \rho_{1} + \rho_{2} \pm 2\sqrt{\rho_{1}\rho_{2}} |\mu|$, which when substituted into Eq.(\ref{11}) provides
\begin{equation}
\label{v1}
V = \frac{2\sqrt{\rho_{1}\rho_{2}} |\mu|}{\rho_{1} + \rho_{2}} =\frac{2r_{1}r_{2}|\rho_{12} + \rho_{34}|}{r_{2}^2(\rho_{11} + \rho_{33}) + r_{1}^2(\rho_{22} + \rho_{44})}.
\end{equation}

At this stage, if we consider the visibility in the region close to the central point of the detection screen, and that the distance between the mask $\mathcal{A}$ and the screen $\mathcal{B}$ is much bigger than the distance between the slits, we can assume $r_{1}^2\approx r_{2}^2\approx r_{1}r_{2}$. Then, using this approximation and the equality $Tr[\hat{\rho}]  = 1$, Eq. (\ref{v1}) reduces to
\begin{equation}
\label{20}
V = 2 |\rho_{12} + \rho_{34}|.
\end{equation}
This relation is our final result for the visibility of interference fringes, which considers the polarization degree of freedom. Note that this is given only in terms of the elements of the coherence-polarization matrix, $\hat{\rho}$.

Now we proceed to the derivation of the quantifier of distinguishability. Here, two physical elements are capable of providing information about which path the photon take in the apparatus. Namely, the path probability and the polarization state in each path. The polarization state in path $j$ (with $j = 1,2$) can be expressed in terms of the respective Stokes parameters as follows \cite{james}:
\begin{eqnarray} \label{21}
\hat{\rho}_{j} = \frac{1}{2}\sum_{i=0}^{3}\frac{S^{(j)}_{i}}{S^{(j)}_{0}}{\hat\sigma}_{i},
\end{eqnarray}
where $\hat{\sigma}_{i}$ are the Pauli matrices. With this and the results of Eqs. (\ref{s01}) to (\ref{s23}) we obtain that
\begin{eqnarray} \label{22}
\hat{\rho}_{1} = \frac{1}{\rho_{11} + \rho_{33}}\left(\begin{matrix}
\rho_{11}& \rho_{13}\\
\rho_{31}&\rho_{33}
\end{matrix}\right),
\end{eqnarray}
and 
\begin{eqnarray} \label{23}
\hat{\rho}_{2} = \frac{1}{\rho_{22} + \rho_{44}}\left(\begin{matrix}
\rho_{22}& \rho_{24}\\
\rho_{42}&\rho_{44}
\end{matrix}\right).
\end{eqnarray}
The difference between the polarization states in each path can be quantified using the trace distance \cite{nielsen},
\begin{equation}
\label{24}
T(\hat{\rho}_{1},\hat{\rho}_{2}) = \frac{1}{2} Tr|\hat{\rho}_{1} - \hat{\rho}_{2}| = \sum_{i}|\lambda_{i}|,
\end{equation}
with $|\hat{A}| = \sqrt{\hat{A}^\dag \hat{A}}$ being the definition of the absolute value of an operator $\hat{A}$, and $\lambda_{i}$ the eigenvalues of $\hat{\rho}_{1} - \hat{\rho}_{2}$. The second equality of Eq. (\ref{24}) is valid because $\hat{\rho}_{1} - \hat{\rho}_{2}$ is a Hermitian operator. The trace distance $T(\hat{\rho}_{1},\hat{\rho}_{2})$ is a metric on the space of density matrices which obeys $0\leq  T(\hat{\rho}_{1},\hat{\rho}_{2}) \leq 1$, with $ T(\hat{\rho}_{1},\hat{\rho}_{2})=0$ iff $\hat{\rho}_{1} = \hat{\rho}_{2}$, and $T(\hat{\rho}_{1},\hat{\rho}_{2}) = 1$ iff $\hat{\rho}_{1}$ and $\hat{\rho}_{2}$ are orthogonal states.

In order to obtain a distinguishability quantifier that also accounts for the path probability, we use the generalized (weighted) trace distance approach, where the weights are given by the path probabilities $I_{1}$ and $I_{2}$ of the photons pass through slit 1 and 2, respectively.
\begin{eqnarray}
\label{25}
 D = Tr|I_{1} \hat{\rho}_{1} - I_{2} \hat{\rho}_{2}| = Tr |\hat{\Delta}|,
\end{eqnarray}
where the operator $\hat{\Delta} = I_{1} \hat{\rho}_{1} - I_{2} \hat{\rho}_{2}$ is the so-called Helstrom matrix \cite{helstrom}. This generalized definition of the trace distance has recently been used in measures of quantum non-Markovianity \cite{chru,amato}. In terms of the elements of $\hat{\rho}$, we have that $I_{1} = S^{(1)}_{0} = \rho_{11} + \rho_{33}$ and $I_{2} = S^{(2)}_{0} = \rho_{22} + \rho_{44}$.

With this, we find that
\begin{equation} \label{27}
D = Tr \left|\left(\begin{matrix}
    (\rho_{11} - \rho_{22})& (\rho_{13} - \rho_{24})\\
    (\rho_{31} - \rho_{42})& (\rho_{33} - \rho_{44})\\
    \end{matrix}\right)\right|.
\end{equation}
The eigenvalues of this matrix are given by
\begin{eqnarray}\label{28}
\lambda_{1} = \frac{\alpha + \sqrt{\alpha^{2} + 4(|\beta|^2 - \kappa)}}{2},
\end{eqnarray}
\begin{eqnarray}\label{29}
\lambda_{2} = \frac{\alpha - \sqrt{\alpha^{2} + 4(|\beta|^2 - \kappa)}}{2},
\end{eqnarray}
with $\alpha = \rho_{11} - \rho_{22} + \rho_{33} - \rho_{44}$, $\beta = \rho_{13} -  \rho_{24}$ and $\kappa = (\rho_{11} - \rho_{22})(\rho_{33} - \rho_{44})$. By using that $\rho_{mn} \rho_{nm}^{*} = |\rho_{nm}|^{2}$, and the result of the second equality of Eq. (\ref{24}), we obtain 
\begin{eqnarray}
\label{30}
D = \frac{1}{2}\{|\alpha + \gamma| + |\alpha - \gamma| \},
\end{eqnarray}
where $\gamma = \sqrt{\alpha^{2} + 4(|\beta|^{2} - \kappa)}$. This is our final expression for the distinguishability given only in terms of the elements of the coherence-polarization matrix $\hat{\rho}$, so that both the influence of the path probability and the polarization state are considered. Note that if we use the triangle inequality $|z+w| \leq |z|+|w|$, with $z=  \alpha + \gamma$ and $w=\alpha - \gamma$, Eq. (\ref{30}) allows to write $D \geq |\alpha|$. This means that $D \geq |I_{1} - I_{2}|$, where the quantity in the right-hand side of this inequality is the definition of distinguishability (predictability) given in Refs. \cite{greenberger,englert}, in which only the influence of the path probability is considered. Naturally, our result is lower bounded by this quantity because we also included the influence of polarization.

\section{Complementarity Relation}

In this section we show that, similar to the EGY inequality shown in Eq.~(\ref{1}), the quantifiers $V$ and $D$ derived in the previous section satisfy a complementarity relation of the form $D^2+V^2 \leq 1$. To start with, we have that the square of the visibility quantifier of Eq.~(\ref{20}) provides
\begin{eqnarray}
\label{4.1}
V^2  =  4\{|\rho_{12}|^2 + |\rho_{34}|^2 + 2 Re(\rho_{12}\rho_{43})\}.
\end{eqnarray}
Here, if we use the property $|\rho_{nm}|^2 \leq \rho_{nn}\rho_{mm}$ \cite{neumann}, we can write the inequality
\begin{eqnarray}
\label{4.2}
V^2  \leq  4\{\rho_{11}\rho_{22} + \rho_{33}\rho_{44} + 2 Re(\rho_{12}\rho_{43})\}.
\end{eqnarray}

Let us now turn our attention to the square of the distinguishability quantifier of Eq.~(\ref{30}), which can be written as
\begin{eqnarray}
\label{4.4}
D^2 = \frac{1}{2}\{\alpha^2 + \gamma^2 + |\alpha^2 - \gamma^2|\}.
\end{eqnarray}
By expressing the variables $\beta$ and $\kappa$ in this equation, we have that
\begin{eqnarray}
\label{4.5}
D^2 = \frac{1}{2}\{2\alpha^2 + 4(|\beta|^{2} - \kappa) + |4(|\beta|^{2} - \kappa)|\}.
\end{eqnarray}
Now, if we use the fact that for any real $x$ we have: i) $x+|x|=0$, if $x<0$; and ii) $x+|x|=2x$, if $x>0$, we can write the following inequality
\begin{eqnarray}
\label{4.6}
D^2 \leq \alpha^2 + 4(|\beta|^2 - \kappa).
\end{eqnarray}
In parallel, one can verify that
\begin{eqnarray}
\label{4.7}
|\beta|^2 = |\rho_{13}|^2 + |\rho_{24}|^2 - 2 Re(\rho_{13}\rho_{42}),
\end{eqnarray}
from which follows that
\begin{eqnarray}
\label{4.8}
|\beta|^2 \leq \rho_{11}\rho_{33} + \rho_{22}\rho_{44} - 2 Re(\rho_{13}\rho_{42}).
\end{eqnarray}

At this stage, if we substitute the expressions of $\alpha^2$, $\kappa$ and the right-hand side of the above inequality in place of $|\beta|^2$ into the inequality presented in (\ref{4.6}), we obtain that 
\begin{eqnarray}
\label{4.9}
D^2 & \leq &  \rho_{11}^2 + \rho_{22}^2 + \rho_{33}^2 + \rho_{44}^2 - 2\rho_{11}\rho_{22} + 2\rho_{11}\rho_{33} \\ \nonumber
 &-&  2\rho_{11}\rho_{44} -  2\rho_{22}\rho_{33}
 + 2\rho_{22}\rho_{44}
 - 2\rho_{33}\rho_{44} \\ \nonumber
 &+& 4 \{[\rho_{11}\rho_{33} + \rho_{22}\rho_{44} - 2 Re(\rho_{13}\rho_{42})] \\ \nonumber
 &-& (\rho_{11}\rho_{33} + \rho_{22}\rho_{44} - \rho_{11}\rho_{44} - \rho_{22}\rho_{33})\}.
\end{eqnarray}
Now, by summing the inequalities (\ref{4.2}) and (\ref{4.9}), we verify that
\begin{eqnarray}
\label{4.10}
D^2 + V^2 & \leq &  \rho_{11}^2 + \rho_{22}^2 + \rho_{33}^2 + \rho_{44}^2  \\ \nonumber
 &+& 2\rho_{11}\rho_{22} +  2\rho_{11}\rho_{33} +  2\rho_{11}\rho_{44}
 \\ \nonumber
 &+& 2\rho_{22}\rho_{33} +  2\rho_{22}\rho_{44} +  2\rho_{33}\rho_{44},
\end{eqnarray}
where we used the fact that $\rho_{12}\rho_{43} = \rho_{13}\rho_{42}$ \cite{neumann}. The above inequality can be rearranged as
\begin{eqnarray}
\label{4.11}
D^2 + V^2  &\leq& (\rho_{11} + \rho_{22})^2 + (\rho_{33} + \rho_{44})^2 \\ \nonumber
&+& 2 (\rho_{11} + \rho_{22})(\rho_{33} + \rho_{44}) \\ \nonumber
&=& (\rho_{11} + \rho_{22} + \rho_{33} + \rho_{44})^2.
\end{eqnarray}
Since $Tr[\hat{\rho}]  = 1$, we finally find that
\begin{eqnarray}
\label{4.11}
D^2 + V^2  \leq 1.
\end{eqnarray}
With this, we prove that our quantifiers of visibility and distinguishability, $V$ and $D$, which takes into account the influences of the path probability and polarization, satisfy a complementarity relation analogous to the EGY inequality. Here, it is worthwhile to point out that, since the polarization degree of freedom is described by an algebra equivalent to that of a spin-1/2 system \cite{sakurai}, all present results can be extended to the case of spin-1/2 particles in a two-path interferometer.     

\section{Examples}

The physical meaning of the quantifiers of visibility and distinguishability derived here can be better understood when applied to some special states. As a first example we consider the pure state of an ensemble of horizontally polarized photons submitted to the double-slit apparatus. The probability amplitudes of the photons passing through slits 1 and 2 are quantified with the real parameter $a$ as follows:
\begin{eqnarray}
\label{31}
\ket{\psi_{1}} = a \ket{H,1} + \sqrt{1- a^2} \ket{H,2},
\end{eqnarray}
 with $a \in [0,1]$. The results for $D$, $V$ and $D^{2}+ V^{2}$ are shown in Fig. 2. As expected, we have particle-like behavior ($D=1$ and $V=0$) for the states $\ket{H,1}$ and $\ket{H,2}$, i.e., when $a=0$ and $a=1$, because in these cases we know the path taken. We have wave-like behavior ($D=0$ and $V=1$) when the state is $1/\sqrt{2}(\ket{H,1} + \ket{H,2})$. Note that $\ket{\psi_{1}}$ satisfies the upper limit of our complementarity relation, Eq.~(\ref{4.11}). In fact, $D^{2}+ V^{2} = 1$ for all values of $a$. As we shall see, this is a  characteristic of pure states.   

\begin{figure}[ht]
\centerline{\includegraphics[width=6.5cm]{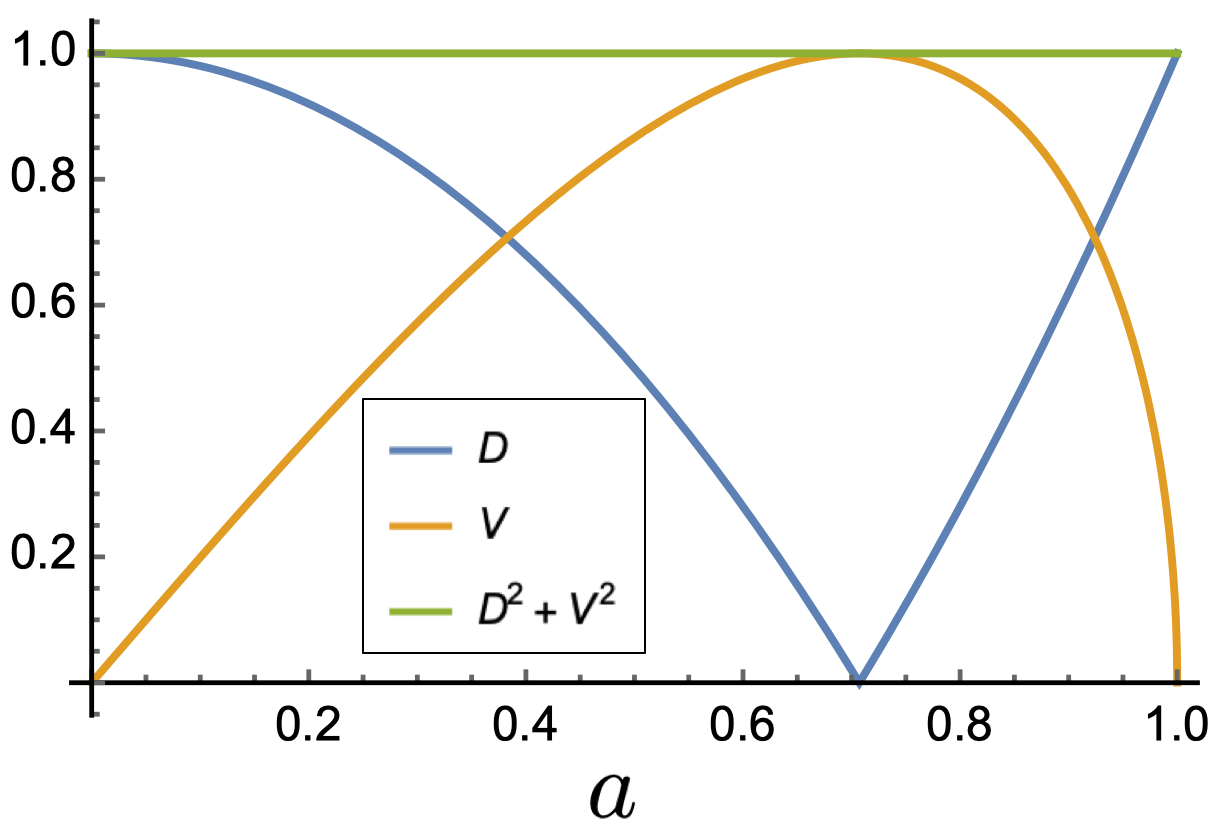}}
\caption{(Color online) Behavior of the distinguishability $D$, the visibility $V$ and the function $D^{2}+ V^{2}$ for the state $\ket{\psi_{1}} = a \ket{H,1} + \sqrt{1- a^2} \ket{H,2}$, with $a \in [0,1]$. The limit cases are $D=1$ and $V=0$ for $a=0,1$; and $D=0$ and $V=1$ for $a=1/\sqrt{2}$.}
\label{graphs}
\end{figure}

Another interesting example concerns the state
\begin{eqnarray}
\label{32}
\ket{\psi_{2}} = a \ket{H,1} + \sqrt{1- a^2} (b \ket{H,2} + i \sqrt{1- b^2} \ket{V,2}), \nonumber \\
\end{eqnarray}
 with the real parameters $a,b \in [0,1]$. When $b=1$ we have $\ket{\psi_{2}} = \ket{\psi_{1}}$, which is the case shown in Fig. 2. If $b=0$ we have $V=0$ and $D=1$, $\forall a$. This case of complete distinguishability is because the polarization states in the paths are orthogonal independently of $a$, $\ket{\psi_{2}} = a \ket{H,1} + i \sqrt{1- a^2} \ket{V,2})$. The $b = 1/\sqrt{2}$ case is especially interesting, and results the for $D$, $V$ and $D^{2}+ V^{2}$ are shown in Fig. 3. As can be seen, when $a = 1/\sqrt{2}$ there is partial visibility and distinguishability, $D=V=1/\sqrt{2}$. This is the case in which the path probabilities are equal, but there is horizontal polarization in path 1 and right circular polarization in path 2, i.e., the non-orthogonality of the polarization states gives rise to this hybrid behavior of particle and wave. Again, since we are dealing with pure states, the relation $D^{2}+ V^{2} =1$ still holds.     
 
\begin{figure}[ht]
\centerline{\includegraphics[width=6.5cm]{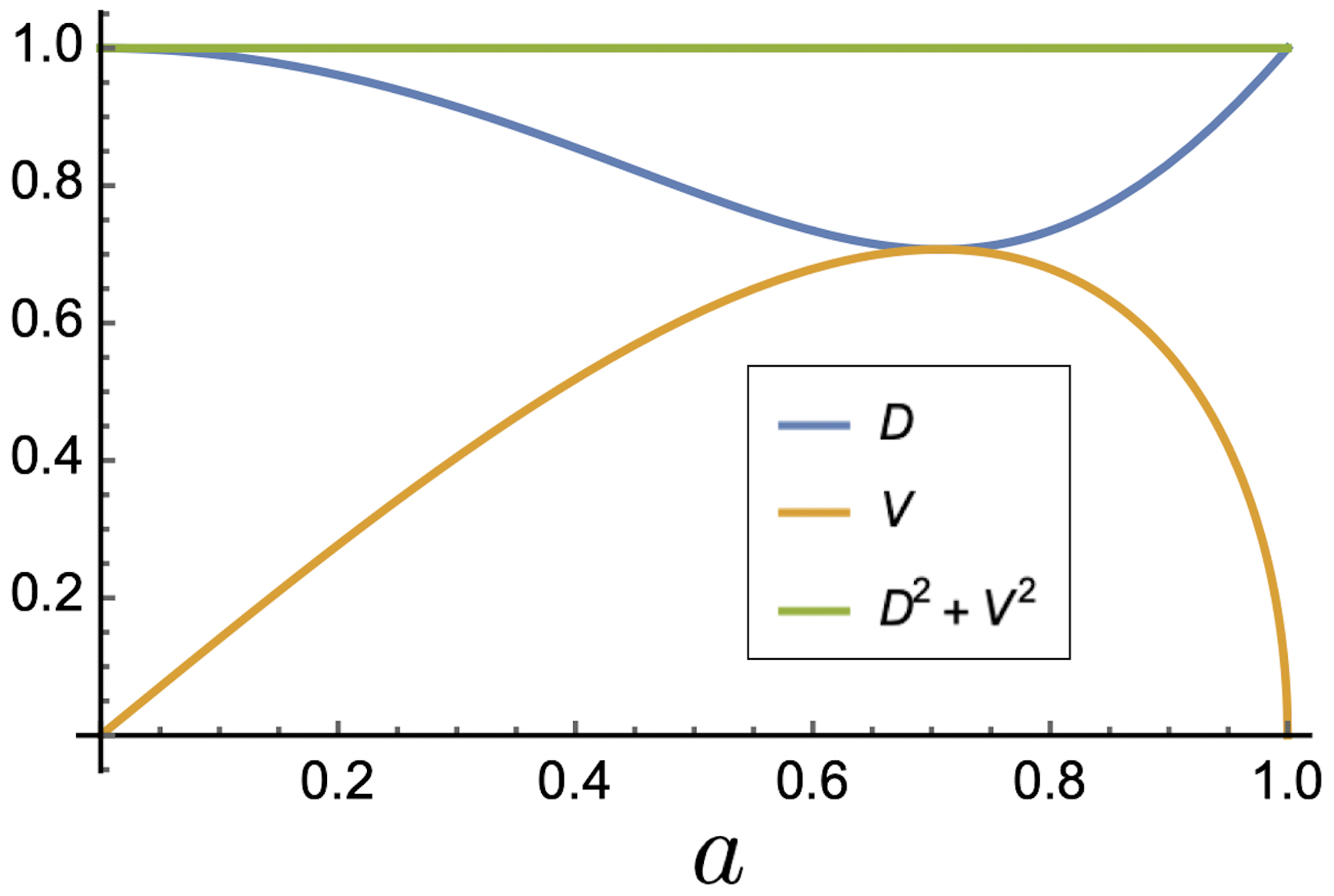}}
\caption{(Color online) Behavior of the distinguishability $D$, the visibility $V$ and the function $D^{2}+ V^{2}$ for the state $\ket{\psi_{2}} = a \ket{H,1} + \sqrt{(1- a^2)/2} (\ket{H,2} + i \ket{V,2})$, with $a \in [0,1]$. The limit cases provide $D=1$ and $V=0$ when $a=0,1$; and $D=V=1/\sqrt{2}$ for $a=1/\sqrt{2}$.}
\label{graphs}
\end{figure}

\begin{figure}[ht]
\centerline{\includegraphics[width=6.5cm]{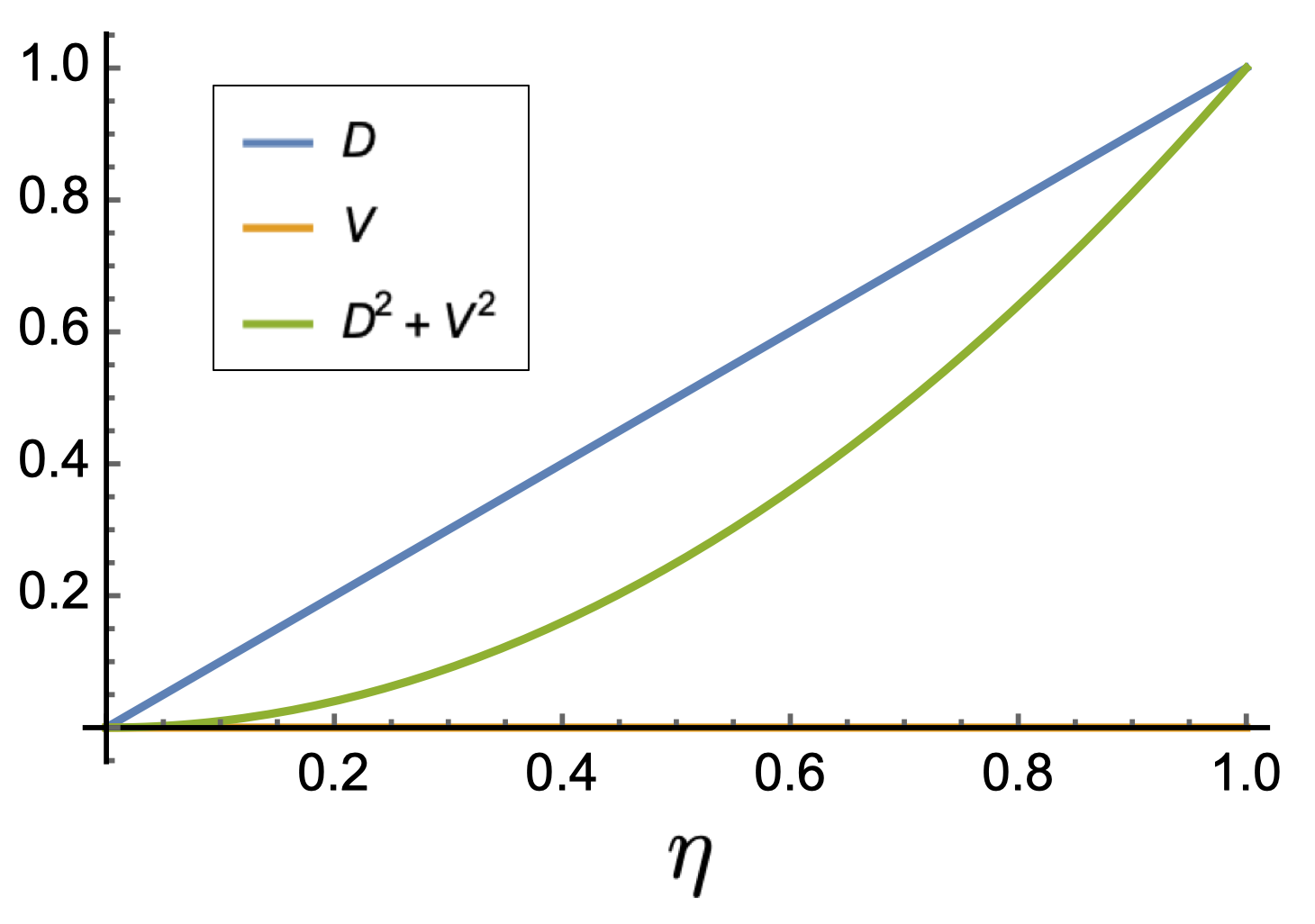}}
\caption{(Color online) Behavior of the distinguishability $D$, the visibility $V$ and the function $D^{2}+ V^{2}$ for the Werner state $\hat{\rho}_{W}$ as a function of the parameter $\eta$. There is no visibility in this case, $V = 0$, and the maximal distinguishability, $D = 1$, only occurs when the state is pure, $\eta = 1$.}
\label{graphs}
\end{figure} 

As a last example, we consider a Werner state involving the path and polarization degrees of freedom in the form:
\begin{eqnarray}
\label{33}
\hat{\rho}_{W} = \eta \ket{\psi^{(-)}}\bra{\psi^{(-)}} + \frac{1 - \eta}{4} \hat{I},
\end{eqnarray}
with $0\leq \eta \leq 1$. The state $\hat{I}$ denotes the 4x4 identity matrix in the basis formed by the vectors $\{ \ket{H,1},\ket{H,2}, \ket{V,1},\ket{V,2} \}$, and $\ket{\psi^{(-)}} = 1/ \sqrt{2}(\ket{H,2}-\ket{V,1})$ is the singlet state, corresponding to a maximally entangled state involving the path and polarization of the photons. As a matter of fact, Werner demonstrated that the state $\hat{\rho}_{W}$ is entangled only if $\eta > 1/3$ \cite{werner}. For this state, the results for $D$, $V$ and $D^{2}+ V^{2}$ are exhibited in Fig. 4. Here, we see that the limit $D^2 + V^2 = 1$ is only attained when the state is pure, $\eta = 1$. For the completely mixed state, $\hat{\rho}_{W} = \hat{I}/4$ (when $\eta =0$), we have complete absence of visibility and distinguishability, $D = V = 0$. That is, the complete lack of phase relation involving the paths and the polarization components of the photons prevent us from obtaining any which-path information and interference pattern. Actually, $V = 0$ for all values of $\eta$. This is because the state $\ket{\psi^{(-)}}$ illuminates the slits with orthogonal polarizations, therefore, making impossible the formation of interference.

\section{Behavior of the distinguishability and visibility in open system dynamics}

An interesting application of the distinguishability and visibility quantifiers in the context of the wave-particle duality is the study of open quantum dynamics. The scenario in question permits describing the action of environments capable of causing decoherence, depolarization, and scattering events to the photons (system) \cite{joos, schlosshauer,zurek}, before they pass through the slits, and investigate how the quantifiers $V$ and $D$ can provide information about the system-environment interaction. To depict this situation, we shall use the operator-sum representation \cite{nielsen}. It is assumed that the system-environment input state is an uncorrelated
state in the form 
\begin{equation}
\label{38}
\hat{\rho}(0) = \hat{\rho}_{S}(0)\otimes\hat{\rho}_{E}(0),
    \end{equation}
where $\hat{\rho}_{S}$ is the operator that represents the coherence-polarization state of the photons,  according to Eq. (\ref{2}), and $\hat{\rho}_{E}$ is the environment state, which is considered to be in a general diagonal form, $\hat{\rho}_{E} = \sum_{i} q_{i}|E_{i}\rangle \langle E_{i}|$, as is the case of a thermal state, for example \cite{bernardo2,bernardo3}. In this framework, one considers that system and environment evolve together under a joint unitary operation $\hat{U}_{SE}$, so that the reduced state of the system evolves according to the relation
\begin{eqnarray}\label{40}
\hat{\rho}_{S}(t) = Tr_{E}\left\{\hat{U}_{SE}\left[\hat{\rho}_{S}\otimes\sum_{i}q_{i}|E_{i}\rangle \langle{E_{i}}|\right]\hat{U}^{\dag}_{SE}\right\},
\end{eqnarray}
where $Tr_{E}$ denotes trace over the states of the environment. This relation can be rewritten as
\begin{eqnarray}
\label{41}
\hat{\rho}_{S}(t) = \sum_{ij} \hat{K}_{ij}\hat{\rho}_{S}\hat{K}_{ij}^{\dag},
\end{eqnarray}
where $\hat{K}_{ij}\equiv \sqrt{q_{i}}\langle{E_{j}}|\hat{U}_{SE}|E_{i}\rangle $ are the so-called Kraus operators, which satisfy the relation  $\sum_{i}\hat{K}_{ij}^\dag \hat{K}_{ij} = \hat{I}$ \cite{nielsen,preskill}. 

To proceed further, we start studying an example of system-environment evolution that only involves decoherence, first presented in Ref. \cite{bernardo}. We suppose that they interact via the following unitary map:  
\begin{equation}
\label{49}
|H,1\rangle|E_{0}\rangle \longrightarrow \sqrt{1-P}|H,1\rangle |E_{0}\rangle + \sqrt{P}|H,1\rangle|E_{1}\rangle, \end{equation}
\begin{equation}
|H,2\rangle|E_{0}\rangle \longrightarrow \sqrt{1-P}|H,2\rangle |E_{0}\rangle + \sqrt{P}|H,2\rangle|E_{2}\rangle, \end{equation}
\begin{equation}
| V,1\rangle|E_{0}\rangle \longrightarrow\sqrt{1-P}|V,1\rangle |E_{0}\rangle + \sqrt{P}|V,1\rangle|E_{1}\rangle, 
\end{equation}
\begin{equation}
\label{52}
| V,2\rangle|E_{0}\rangle \longrightarrow \sqrt{1-P}|V,2\rangle |E_{0}\rangle + \sqrt{P}|V,2\rangle|E_{2}\rangle.
\end{equation}
These relations describe the case of an environment made up of small refractive particles uniformly distributed in space that cause random phase shifts to the photons. The state of the environment is initially $\ket{E_{0}}$, which can change to one of the orthogonal states $\{\ket{E_{1}},\ket{E_{2}}\}$, depending on which path is taken by the photon. The parameter $P$ is the probability of occurring an interaction between a photon and an environment constituent during a given time interval $\Delta t$.  

The set of Kraus operators $\hat{K}_{j} =\langle{E_{j}}|\hat{U}_{SE}|{E_{0}}\rangle$ that rules the dynamics of the photons is
\begin{eqnarray}
 \hat{K}_{0} &= &\sqrt{1-P}(|H,1\rangle\langle{H,1}|  + |H,2\rangle\langle{H,2}| \nonumber\\
& & + |V,1\rangle\langle{V,1}| + |V,2\rangle\langle{V,2}|),
\end{eqnarray}
\begin{equation}
\hat{K}_{1} = \sqrt{P} (|H,1\rangle\langle{H,1}| + |V,1\rangle\langle{V,1}| ), 
\end{equation}
\begin{equation}
\hat{K}_{2} = \sqrt{P} (|H,2\rangle\langle{H,2}| + |V,2\rangle\langle{V,2}|).
\end{equation}
The substitution of these results in Eq. (\ref{41}) allows us to express the state evolution of the system as a function of $P$. To describe this evolution as a function of time we assume that the probability of an interaction event per unit time is $\Gamma$, so that $P = \Gamma  \Delta t \ll 1$ during a short time interval $\Delta t$. With this, the state evolution of the system after a time
$t = n \Delta t$ is a result of the application of the map $n$ times successively. This is an assumption of {\it Markovian} system-environment dynamics, in which the time evolution of the system at the present time has {\it no memory} of the
past \cite{breuer2,vega,li}. This permits us to write the probabilistic relation $(1-p) \rightarrow (1-p)^{n} = \lim_{n \to \infty} \left(1-\frac{\Gamma t}{n} \right)^n = e
^{- \Gamma t}$, where $\Delta t \rightarrow 0$ was assumed.
Taking all these points into consideration, the time evolution of the photons becomes
\begin{eqnarray}\label{43}
\hat{\rho}_{S}(t) = 
\left (
\begin{array}{cccc}
\rho_{11} & \gamma\rho_{12}& \rho_{13} & \gamma\rho_{14}\\
\gamma \rho_{21}& \rho_{22} & \gamma\rho_{23} & \rho_{24} \\
 \rho_{31} & \gamma\rho_{32} &  \rho_{33} & \gamma\rho_{34} \\
\gamma\rho_{41}& \rho_{42} & \gamma \rho_{43} & \rho_{44} 
\end{array}
\right),
\end{eqnarray}
where $\gamma = e^{-\Gamma t}$.

\begin{figure}[ht]
\centerline{\includegraphics[width=6.5cm]{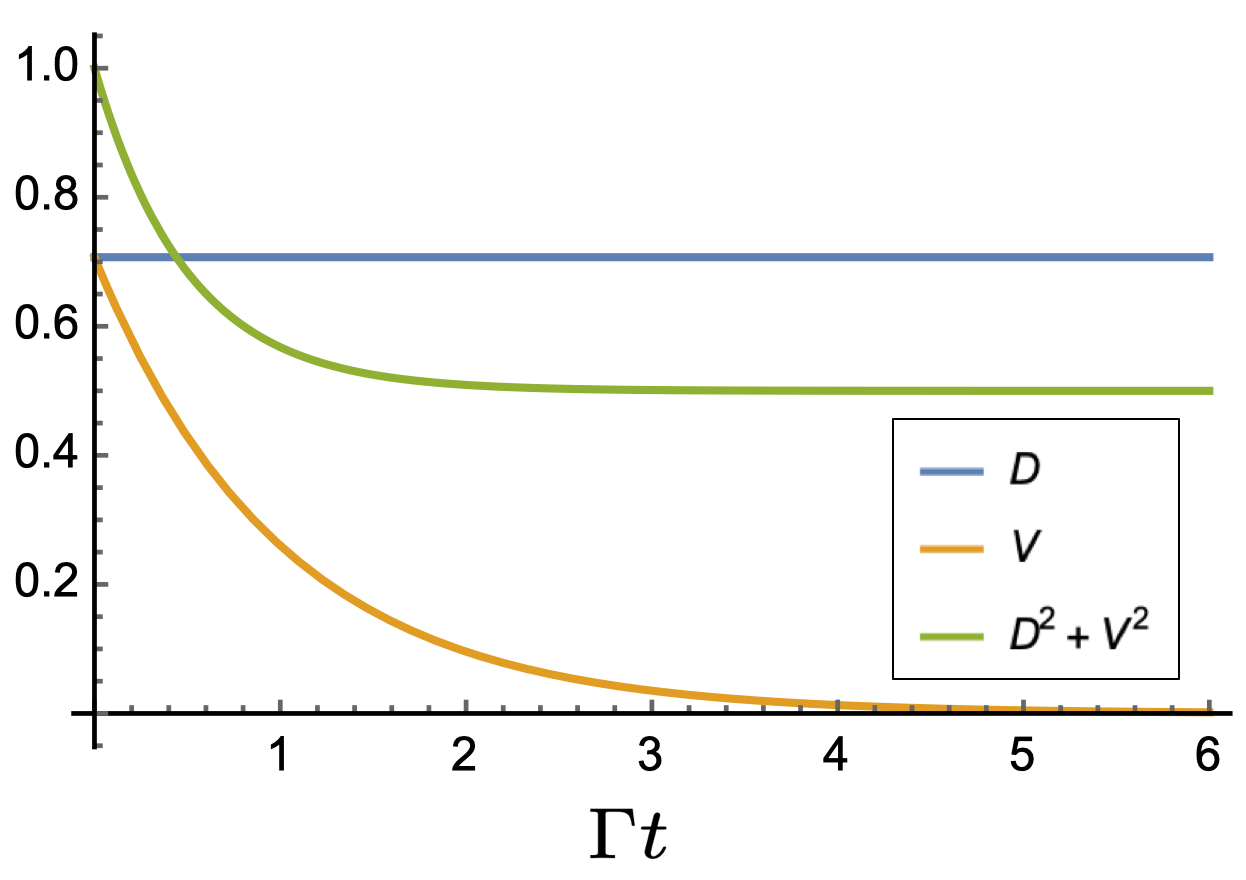}}
\caption{(Color online) Time evolution of the distinguishability $D$, visibility $V$ and the function $D^{2}+ V^{2}$ for the state $\ket{\psi}$ of Eq. (\ref{state}) under the environment action determined by Eqs. (\ref{49}) to (\ref{52}). $D$ is found to be constant and $V$ presents an exponential decay dynamics.}
\end{figure}

Now, to illustrate the time evolution of $D$ and $V$ under the action of this interaction, it is interesting to choose an initial state in which these two quantities are non-zero at the initial time. Here, we assume an ensemble of photons prepared in an equal superposition of horizontal polarization in path 1, $\ket{H,1}$, and diagonal polarization in path 2, $\ket{D,2} = 1/\sqrt{2}(\ket{H,2}+\ket{V,2})$. Such a state is given by
\begin{equation}
\label{state}
\ket{\psi} = \frac{1}{\sqrt{2}} \ket{H,1} + \frac{1}{2}(\ket{H,2}+\ket{V,2}).
\end{equation}
This is a pure state with partial distinguishability and visibility. The evolution of these quantities are shown in Fig. 5. The results are in agreement with what we expect from Eqs.~(\ref{49}) to (\ref{52}). Indeed, we obtained an exponential decay for the visibility, which is a consequence of the random phase shifts continuously imparted to the photons by the environment, which causes decoherence. In turn, the distinguishability has a constant value, $D=1/\sqrt{2}$. This is because the path probabilities do not change with time, since the evolution map does not predict photon scattering events, and the polarization states associated with each path are preserved.

\begin{figure}[ht]
\centerline{\includegraphics[width=6.5cm]{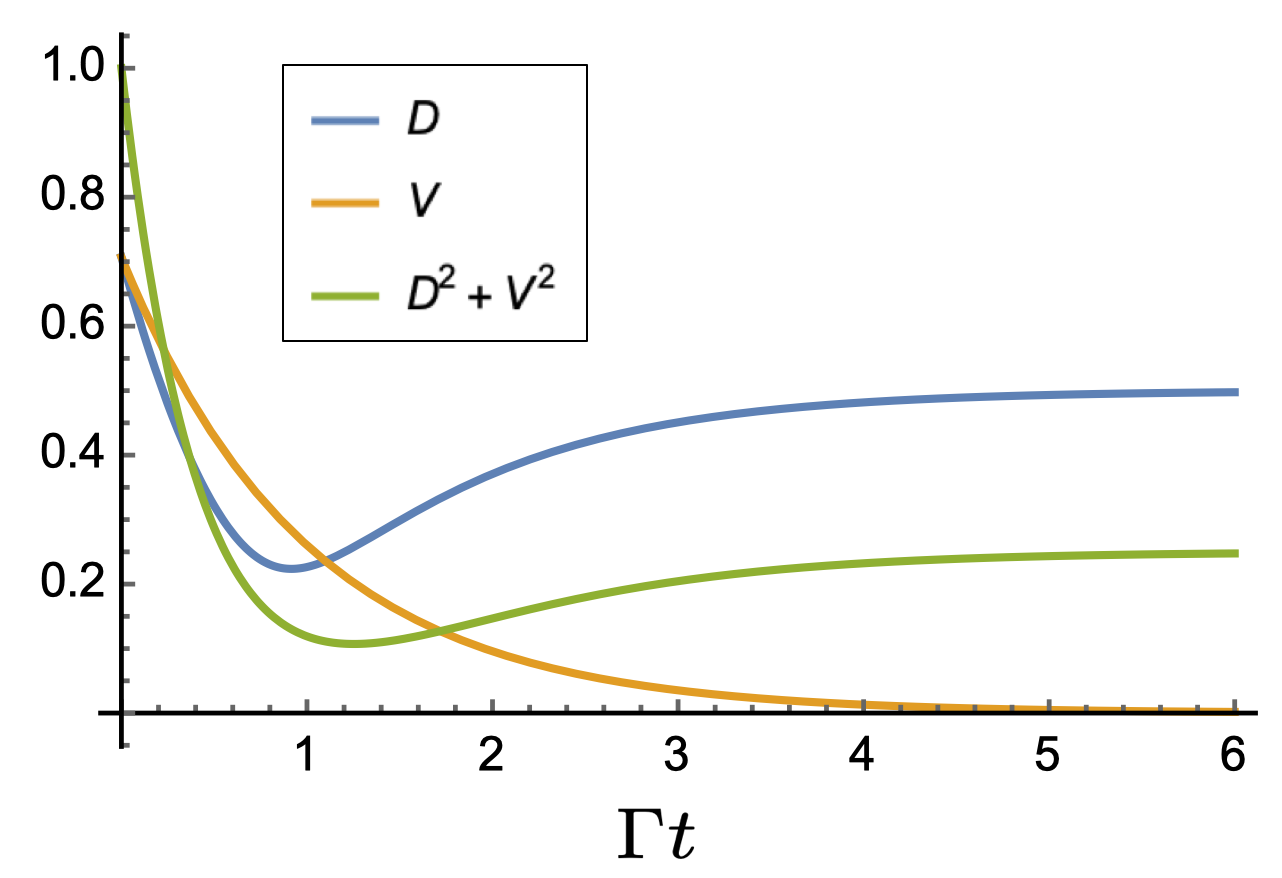}}
\caption{(Color online) Time evolution of the distinguishability $D$, the visibility $V$ and the function $D^{2}+ V^{2}$ for the state $\ket{\psi}$ of Eq. (\ref{state}) under the environment influence determined by Eqs. (\ref{58}) to (\ref{61}). $V$ presents an exponential decay dynamics, whereas $D$ exhibits an interesting down-and-up behavior.}
\label{graphss}
\end{figure}

Our second example addresses a system-environment interaction that
involves both decoherence and photon scattering. The dynamics is described by the following unitary map: 
\begin{equation}
\label{58}
|H,1\rangle|E_{0}\rangle \longrightarrow \sqrt{1-P}|H,1\rangle |E_{0}\rangle + \sqrt{P}|H,2\rangle|E_{1}\rangle, \end{equation}
\begin{equation}
|H,2\rangle|E_{0}\rangle \longrightarrow \sqrt{1-P}|H,2\rangle |E_{0}\rangle + \sqrt{P}|H,1\rangle|E_{2}\rangle, \end{equation}
\begin{equation}
|V,1\rangle|E_{0}\rangle \longrightarrow\sqrt{1-P}|V,1\rangle |E_{0}\rangle + \sqrt{P}|V,2\rangle|E_{3}\rangle, 
\end{equation}
\begin{equation}
\label{61}
|V,2\rangle|E_{0}\rangle \longrightarrow \sqrt{1-P}|V,2\rangle |E_{0}\rangle + \sqrt{P}|V,1\rangle|E_{4}\rangle.
\end{equation}
Here, the environment causes random polarization-dependent phase shifts and scattering events to the photons. In practice, this could take place in an environment composed of small birefringent particles capable of deviate the photon trajectories \cite{wai}. The relations above tell us that photons in a given path are scattered to the opposite path if an interaction occurs. The state of the environment, which starts out in $\ket{E_{0}}$, will change to one of the orthogonal states $\{\ket{E_{1}},\ket{E_{2}},\ket{E_{3}},\ket{E_{4}}\}$, depending on the path and polarization of the photons. Again, the parameter $P$ is the interaction probability during a time interval $\Delta t$.

With this, the Kraus operators $\hat{K}_{j} =\langle{E_{j}}|\hat{U}_{SE}|{E_{0}}\rangle$ that dictate the photon state evolution are
\begin{eqnarray}
\hat{K}_{0} &= &\sqrt{1-P}(|H,1\rangle\langle{H,1}|  + |H,2\rangle\langle{H,2}| \nonumber\\
& & + |V,1\rangle\langle{V,1}| + |V,2\rangle\langle{V,2}|),
\end{eqnarray}
\begin{eqnarray}
\hat{K}_{1} = \sqrt{P} (|H,2\rangle\langle{H,1}|), 
\end{eqnarray}
\begin{eqnarray}
\hat{K}_{2} = \sqrt{P} (|H,1\rangle\langle{H,2}|), 
\end{eqnarray}
\begin{eqnarray}
\hat{K}_{3} = \sqrt{P} (|V,2\rangle \langle{V,1}|), 
\end{eqnarray}
\begin{eqnarray}
\hat{K}_{4} = \sqrt{P} (|V,1\rangle \langle{V,2}|).
\end{eqnarray}

Following the same procedure of the previous example, and assuming again a Markovian system-environment interaction, the time evolution of the state of the photons can be written as
\begin{widetext}
\begin{eqnarray}
\label{43}
\hat{\rho}_{S}(t) =
\left (
\begin{array}{cccc}
\gamma \rho_{11} + \epsilon\rho_{22}& \gamma\rho_{12}& \gamma\rho_{13} & \gamma\rho_{14}\\
\gamma \rho_{21}& \gamma\rho_{22} + \epsilon\rho_{11}& \gamma\rho_{23} & \gamma\rho_{24} \\
\gamma \rho_{31} & \gamma\rho_{32} & \gamma \rho_{33} + \epsilon\rho_{44}& \gamma\rho_{34} \\
\gamma\rho_{41}& \gamma \rho_{42} & \gamma \rho_{43} & \gamma \rho_{44} + \epsilon\rho_{33}
\end{array}
\right), 
\end{eqnarray}
\end{widetext}
where $\gamma = e^{-\Gamma t}$ and $\epsilon = (1-e^{-\Gamma t})$.

For the sake of comparison, here we also use the same initial state of the previous example, shown in Eq. (\ref{state}), to illustrate different aspects present in the open quantum dynamics described by Eqs. (\ref{58}) to (\ref{61}). In this case, the evolution of the quantifiers $D$ and $V$, along with the complementarity function $D^2 + V^2$, are shown in Fig. 6. As before, an exponential decay of the visibility is observed. However, an interesting down-and-up behavior takes place for the distinguishability, which is not a characteristic usually observed in the dynamics of Markovian processes \cite{breuer,liu}. In the long-time limit, interference fringes are no longer observed, $V \rightarrow 0$, but the distinguishability is non-zero, $D \rightarrow 1/2$. This is because the long-time state, despite having the same path probabilities, is unpolarized  in path 1 and horizontally-polarized in path 2. This results in partial which-path information encoded in the polarization degree of freedom.

\section{Conclusion}

In conclusion, we have presented an information-theoretic framework to study the complementarity principle in the double-slit scenario for photons with arbitrary path and polarization states. We derived expressions for the path distinguishability $D$ and interference visibility $V$, explicitly in terms of the elements of the coherence-polarization density matrix, which account for the combined effect of coherence and polarization. The generality of the obtained expressions makes our approach different from previous studies of the influence of internal degrees of freedom on wave-particle duality \cite{zela,eberly,banaszek}. Remarkably, despite the generality, we demonstrated that $D$ and $V$ satisfy a complementarity relation analogous to the EGY inequality, $D^2 + V^2 \leq 1$. The equality in this relation was observed to be valid only when the coherence-polarization state is pure.

As a further matter, the simplicity of the present framework allowed us to explore the influence of different environment actions on the photon states by studying the dynamics of $D$ and $V$, where the operator-sum representation was used. This investigation may reveal important aspects of system-environment interactions. For example, it was shown a Markovian dynamics in which the path distinguishability exhibited an unexpected down-and-up behavior with time. As such, we believe that the present work also opens up a new perspective in using optical systems to inspect open quantum dynamics. In future works, we intend to test the present viewpoint in non-Markovian scenarios.

\section*{ACKNOWLEDGMENTS}
        
The authors acknowledge support from Coordena{\c c}{\~a}o de Aperfei{\c c}oamento
de Pessoal de N{\'i}vel Superior (CAPES, Finance Code 001), and Conselho Nacional de Desenvolvimento Cient{\'i}fico e Tecnol{\'o}gico (CNPq). BLB acknowledges support from (CNPq, Grant No. 303451/2019-0), Pronex/Fapesq-PB/CNPq, Grant No. 0016/2019, and PROPESQ/PRPG/UFPB (Project code PIA13177-2020).

\end{document}